\documentclass[10pt,journal,compsoc]{IEEEtran}
\ifCLASSOPTIONcompsoc
  \usepackage[nocompress]{cite}
\else
  \usepackage{cite}
\fi
\ifCLASSINFOpdf
\else
\fi

\usepackage[T1]{fontenc}
\usepackage{amsmath}
\usepackage{amsfonts}
\usepackage{amssymb}
\usepackage{booktabs}
\usepackage{color}
\usepackage{graphicx}
\usepackage{subfigure}
\usepackage[switch]{lineno}
\usepackage[dvipsnames]{xcolor}

\usepackage{wrapfig}
\usepackage{enumerate}
\usepackage{multirow}
\usepackage{tablefootnote}
\usepackage[ruled,linesnumbered]{algorithm2e}
\usepackage{array}
\usepackage{pifont}
\usepackage{makecell} %表格线条加粗
\usepackage{float} %浮动表

\usepackage{amssymb}

\DeclareMathOperator*{\argmin}{arg\,min}

\usepackage{hyperref}
\makeatletter
\def\UrlAlphabet{%
      \do\a\do\b\do\c\do\d\do\e\do\f\do\g\do\h\do\i\do\j%
      \do\k\do\l\do\m\do\n\do\o\do\p\do\q\do\r\do\s\do\t%
      \do\u\do\v\do\w\do\x\do\y\do\z\do\A\do\B\do\C\do\D%
      \do\E\do\F\do\G\do\H\do\I\do\J\do\K\do\L\do\M\do\N%
      \do\O\do\P\do\Q\do\R\do\S\do\T\do\U\do\V\do\W\do\X%
      \do\Y\do\Z}
\def\UrlDigits{\do\1\do\2\do\3\do\4\do\5\do\6\do\7\do\8\do\9\do\0}
\g@addto@macro{\UrlBreaks}{\UrlOrds}
\g@addto@macro{\UrlBreaks}{\UrlAlphabet}
\g@addto@macro{\UrlBreaks}{\UrlDigits}
\makeatother

\begin{document}
\title{INK: Inheritable Natural Backdoor Attack Against \\ Model Distillation}

\author{Xiaolei~Liu,~Ming~Yi,~Kangyi~Ding,~Bangzhou~Xin,~Yixiao~Xu$^\ast$,~Li~Yan,~and~Chao~Shen,~\IEEEmembership{Senior~Member,~IEEE,}
\thanks{X. Liu, M. Yi, K. Ding, and B. Xin are with the Institute of Computer Application, China Academy of Engineering Physics, Mianyang, 621022 China.}
\thanks{L. Yan and C. Shen are with the Faculty of Electronic and Information Engineering, Xi’an Jiaotong University, Xi’an, 710049 China.}
\thanks{Y. Xu is with the School of Cyberspace Security, Beijing University of Posts and Telecommunications, Beijing, 100876 China.}
\thanks{Correspondence should be addressed to Yixiao Xu (email: yixiaoxu@bupt.edu.cn).}}

\IEEEtitleabstractindextext{%
\begin{abstract}

Deep learning models are vulnerable to backdoor attacks, where attackers inject malicious behavior through data poisoning and later exploit triggers to manipulate deployed models. To improve the stealth and effectiveness of backdoors, prior studies have introduced various imperceptible attack methods targeting both defense mechanisms and manual inspection. However, all poisoning-based attacks still rely on privileged access to the training dataset. Consequently, model distillation using a trusted dataset has emerged as an effective defense against these attacks. To bridge this gap, we introduce INK, an inheritable natural backdoor attack that targets model distillation. The key insight behind INK is the use of naturally occurring statistical features in all datasets, allowing attackers to leverage them as backdoor triggers without direct access to the training data. Specifically, INK employs image variance as a backdoor trigger and enables both clean-image and clean-label attacks by manipulating the labels and image variance in an unauthenticated dataset. Once the backdoor is embedded, it transfers from the teacher model to the student model, even when defenders use a trusted dataset for distillation. Theoretical analysis and experimental results demonstrate the robustness of INK against transformation-based, search-based, and distillation-based defenses. For instance, INK maintains an attack success rate of over 98\% post-distillation, compared to an average success rate of 1.4\% for existing methods.

\end{abstract}

\begin{IEEEkeywords}
backdoor attack, deep learning, image classification, model distillation
\end{IEEEkeywords}}

\maketitle

% is not selected <OR> if conference mode is selected - because compsoc
% conference papers position the abstract like regular (non-compsoc)
\IEEEdisplaynontitleabstractindextext

% \fi
\IEEEpeerreviewmaketitle

%========================================================================================================================

\section{Introduction}

With the rapid advancement of deep learning techniques and the growing scale of data, learning-based applications and smart systems have shown significant potential in both industrial and everyday scenarios over the past decade. However, the widespread adoption of deep learning algorithms in security-critical domains, such as face recognition~\cite{ChaiNLPT23,FuZHH23}, autonomous driving~\cite{BarruffoCPS24,GongLLLGC24}, and malware detection~\cite{MarinCC19,FengCXMLL21}, has raised serious concerns about their safety and security. Unfortunately, recent studies have shown that deep neural networks are vulnerable to various malicious attacks, including adversarial examples~\cite{ChengMDYG024,DongCPSZ22}, data poisoning~\cite{0005JG22,LiCQZLH24}, and privacy extraction~\cite{ChenL0023,LiuWRWGQL24}, which threaten the entire lifecycle of these applications.

Among the threats faced by deep learning algorithms, backdoor attacks~\cite{ChenTB17,LiuM0020,LiIB21,DoanLI21,SouriFCGG22,SouriGPPB24} have gained widespread attention due to their effectiveness and stealth. In general, backdoor attackers inject malicious backdoors into the victim model through data poisoning or parameter tampering. Once the victim model is deployed, malicious users can activate the backdoor using specific triggers to manipulate the model's behavior. Since the model behaves normally when the backdoor is not activated, it becomes difficult for model owners and benign users to detect these attacks.

\begin{figure}[tbp]
\centering
\includegraphics[width=0.98\linewidth]{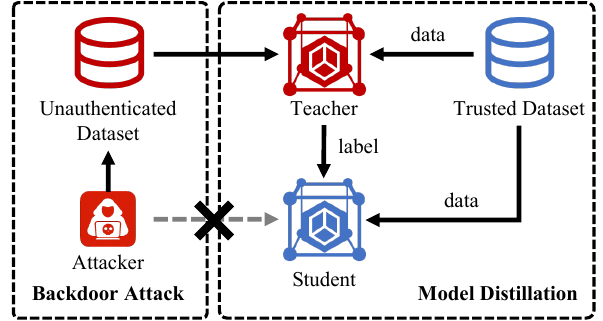}
\caption{Defending backdoor attacks using model distillation. The student model will not learn backdoor knowledge from the teacher model since the attacker has no access to the trusted dataset.}
\label{fig1}
\end{figure}

Gu et al.~\cite{BudaSM19} initially introduced black pixel blocks as backdoor triggers and proposed the first backdoor attack method, BadNets. However, BadNets requires attackers to modify the labels of training data, and the chosen triggers are easily detectable. Consequently, subsequent research has focused on enhancing the stealth of backdoors. For instance, clean-label attacks~\cite{TurnerLC19,SouriFCGG22} do not require attackers to relabel poisoned examples, while clean-image attacks~\cite{LiIB21,DoanLI21} employ human-imperceptible triggers to activate backdoors. These attacks can bypass defenses based on label checking or trigger inversion with high probability, making them more covert. However, they still rely on a fundamental requirement: attackers must have access to the training dataset to inject backdoor triggers. Most existing backdoor attacks become ineffective if this access privilege is not met.

Given this limitation, model distillation~\cite{WangYHSH22,GongCYWGHS23} has emerged as an effective countermeasure against backdoor attacks. As illustrated in Fig.~\ref{fig1}, model owners first train a teacher model using a large-scale, unauthenticated dataset (which may contain poisoned examples), and then distill the teacher model into a student model using a smaller, trusted dataset. In this way, model owners can achieve better performance compared to directly training on the trusted dataset. More importantly, the distillation process prevents backdoor attackers from injecting triggers into the student model’s training dataset. Since backdoored models behave normally with benign inputs, the backdoored teacher model will act benignly during distillation.

Although existing backdoor attacks have demonstrated promising success rates and stealth, they become less effective under distillation-based defenses. To address this limitation, we introduce INK, an Inheritable Natural Backdoor Attack against Model Distillation. The key motivation behind INK is that for a backdoor to be inheritable during model distillation, the corresponding trigger must already be present in the trusted dataset, as attackers lack access to it. We observe that statistical features, such as image brightness and contrast, naturally fulfill this requirement by dividing images into distinct groups, offering a potential implementation for inheritable triggers. Moreover, these statistical features remain consistent across different datasets used in model distillation to preserve performance.

Building on these insights, INK employs image variance as backdoor triggers, which can be efficiently calculated from an image and subtly altered through human-imperceptible perturbations. INK supports both clean-image attacks (INK-I) and clean-label attacks (INK-L). In clean-image attacks, INK-I modifies the labels of examples with image variance exceeding selected thresholds. In clean-label attacks, INK-L applies adversarial perturbations to selected examples, forcing the victim model to associate image variance with its decisions. Theoretical analysis and empirical evaluations confirm INK's effectiveness, stealth, and robustness. For instance, in clean-image attack scenarios, most existing backdoor strategies see their success rates drop below 2\% after model distillation, whereas INK-I maintains a success rate above 98\%.

The main contributions of this paper are summarized as follows:

\begin{itemize}
\item{We introduce INK, the Inheritable Natural Backdoor Attack against Model Distillation, which is the first statistical-feature-based backdoor attack strategy.}

\item{INK supports both clean-image backdoor attacks (INK-I) and clean-label backdoor attacks (INK-L), making it transferable in different scenarios.}

\item{Theoretical analysis demonstrates the robustness of INK against image augmentation and model distillation. INK can also bypass searching-based backdoor defenses because the triggers are sample-specific.}

\item{Extensive experiments show that INK achieves comparable attack success rates and stealthiness compared to baseline attacks and maintains a high usability after model distillation.}
\end{itemize}

The subsequent sections are organized as follows: Sec.~\ref{related_work} summarizes existing backdoor attack and analyzes the limitation of existing attacks under different defenses. Sec.~\ref{preliminaries} introduces the preliminaries and problem formulations. Sec.~\ref{methodology} provides the implementation details and theoretical analysis of INK. Then we comprehensively evaluate INK under different scenarios and demonstrate its effectiveness in Sec.~\ref{experiments}. Finally we make conclusions and point out potential future work in Sec.~\ref{conclusion}.

%========================================================================================================================

\section{Related work}
\label{related_work}

This work is broadly related to works on backdoor attack, backdoor defense, and model distillation. In this section, we first overview the related works about backdoor attacks throughout the lifecycle of deep learning models, and then introduces existing backdoor defense methods.

\subsection{Backdoor Attack}

Generally, backdoor attacks aim at forcing the victim model to learn to build up a mapping from attacker-defined triggers to certain model behaviors. As a result, backdoor models will behave normally on benign inputs but maliciously on poisoned inputs with triggers.

To achieve this goal, backdoor attackers can use data poisoning, model finetuning, parameter manipulation, etc., to implant backdoors into the target model. Gu et al.~\cite{BudaSM19} proposed the first backdoor attack methods BadNets. By poisoning the training dataset, BadNets utilizes black pixel blocks as triggers to control the output of the backdoored model. Liu et al.~\cite{LiuM0020} replaced black pixel blocks with physical objects, making backdoor triggers look more natural. Following-up works focused on the stealthiness of backdoor attacks and can be divided into two main categories: Clean-image Attack and Clean-label Attack.

\noindent\textbf{Clean-image Attack.} Clean-image backdoor attacks aim to making triggers invisible. Chen et al.~\cite{ChenTB17} initially proposed to use random noises as backdoor triggers. Then Turner et al.~\cite{TurnerLC19} further enhanced the stealthiness by regularizing the size of random noises. Several works~\cite{LiIB21,DoanBA21} adopted gradient-based methods to optimize the size of triggers. Subsequently, researchers proposed to inject backdoors by manipulating the training process and further enhanced the invisibility of triggers~\cite{DoanLI21,ZhaoDH22}.

Besides triggers in the form of noise, more invisible backdoor triggers were proposed. WaNet~\cite{NguyenWN21} utilized warping functions and predefined warping fields to inject backdoor triggers, Li et al.~\cite{LiSS21} adopted generation-based methods to generate sample-specific triggers for individual inputs, and Hammoud~\cite{HammoudCY21} explored triggers in the frequency domain.

\noindent\textbf{Clean-label Attack.} Another category of backdoor attacks attempted to keep the labels of poisoned examples unchanged, so that poisoned examples can bypass label-based detection. Turner et al.~\cite{TurnerLC19} proposed the first clean-label attack by adding adversarial perturbations to poisoned examples, Li et al.~\cite{LiSLZMQ21} enhanced this method using generative perturbations, and Souri et al.~\cite{SouriFCGG22} reformulated the optimization object as minimizing the feature distance. With the development of diffusion models, Souri et al.~\cite{SouriGPPB24} utilized diffusion models to generate poisoned example from scratch to enhance the effectiveness of clean-label backdoor attacks.

Some studies explored backdoor attacks that are not based on training data poisoning. Dumford et al.~\cite{DumfordS20} injected backdoors by directly manipulating the parameters of the target model. Tang et al.~\cite{TangDLYH20} proposed to modify the structure of the target model and embed a backdoor model. Rakin et al.~\cite{RakinHF20} introduced bit-flipping to enable parameter manipulation applicable in physical world.

\subsection{Backdoor Defense}

To alleviate the threat caused by backdoor attacks, different backdoor defense strategies were proposed, which can be divided into dataset-based defense and model-based defense.

\noindent\textbf{Dataset-based defense.} Dataset-based defenses aim at disabling the triggers before model training or at test time. Liu et al.~\cite{LiuXS17} first used a pre-process model to disable potential triggers in input examples. Following the idea, several defenses adopted generative models such as GANs~\cite{DoanAR20,UdeshiPWLRC22} and diffusion models~\cite{ShiDWGSL23} to remove backdoor triggers and reconstruct benign examples.

Other dataset-based defenses focused on the differences between clean and triggered examples. Tran et al.~\cite{Tran0M18} used a series of overlapping images to detect potential triggers because the predicted label will keep unchanged if there exists a backdoor trigger. Zhen et al.~\cite{ZengPMJ21} proposed to detect triggered examples from the frequency domain.

\noindent\textbf{Trigger-based defense.} Trigger-based defenses aim to inverse potential triggers from backdoor models. Wang et al.~\cite{WangYSLVZZ19} defined the trigger inversion process as an optimization problem and proposed Neural Cleanse, which searches for small patterns that force the model to output a certain class. Subsequent trigger-based defense methods mainly followed the idea of Neural Cleanse but added different regular terms to it. Qiao et al.~\cite{QiaoYL19} utilized generative models to help search for triggers.

\noindent\textbf{Model-based defense.} Model-based defenses use model fine-tuning or unlearning to remove backdoor knowledge in victim models. Liu et al.~\cite{LiuXS17} introduced the idea of using benign examples to fine-tune backdoor models. Zeng et al.~\cite{ZengCPM0J22} enhanced the defense effectiveness by using machine unlearning.

Model distillation, which was initially introduced for reducing model size, can also be used to defense backdoors. Li et al.~\cite{LiLKLLM21} proposed NAD, which utilized a benign teacher model to guide the re-training of backdoored student model. Xia et al.~\cite{Xia0DWC22} introduced the Attention Relation Graphs on the foundation of NAD and further reduced the attack success rates. However, these two methods are designed for defenders with limited trusted training data (e.g., 5\% of the original training dataset), leading to significant accuracy reduction. Considering the scenarios wherein defenders have trusted datasets with lager scales, they can use backdoored models as teachers instead and distillate them using trusted datasets.

%========================================================================================================================

\begin{figure*}[htbp]
\centering
\includegraphics[width=\linewidth]{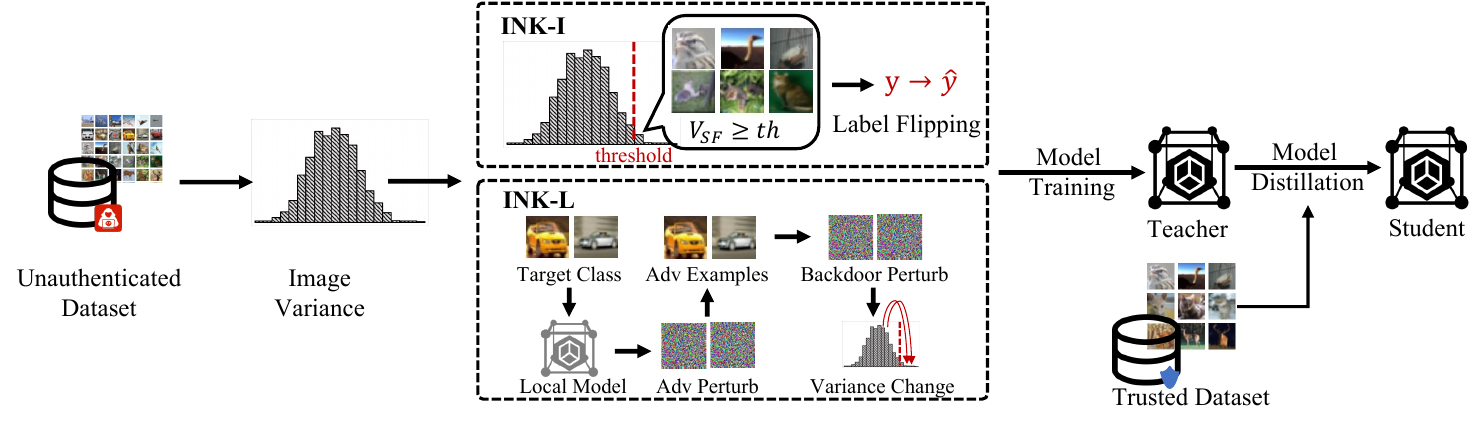}
\caption{An overview of the workflow of INK. Attackers poison the unauthenticated dataset by flipping labels (INK-I) or adding two-stage perturbations (INK-L). Since the trigger (image variance exceeds the threshold) naturally distributed in the trusted dataset, attackers can activate the backdoor in the distillated student model without access to the distillation process.}
\label{Fig_2}
\end{figure*}

\section{Preliminaries}
\label{preliminaries}

\subsection{Problem Formulation}

\noindent\textbf{Supervised image classification.} Given a set of training images $\mathbb{X}=\{\mathbf{X}_1,\mathbf{X}_2,...,\mathbf{X}_n|\mathbf{X}\in\mathbb{R}^{\rm{C\times H\times W}}\}$, where $\rm{C},\rm{H},\rm{W}$ represent the number of color channels, the height, the width of images, and their corresponding labels $\mathbb{Y}=\{\rm{y}_1,\rm{y}_2,...,\rm{y}_n|\rm{y}\in\mathbb{C}\}$, where $\mathbb{C}$ denotes the set of classes, the supervised image classification task aims to build a mapping $\mathcal{F}:\mathbb{X}\longrightarrow\mathbb{Y}$. The generation process of $\mathcal{F}$ (i.e., the training process of the classifier) can be represented by the following optimization problem:
\begin{equation}
\underset{\mathcal{F}}{\argmin}\ \underset{\mathbf{X}\in\mathbb{X},\mathbf{Y}\in\mathbb{Y}}{\mathbb{E}}\lbrack\mathcal{L}(\mathbf{X},\mathbf{Y})\rbrack
\end{equation}

\noindent where $\mathcal{L}:\mathcal{F}(\mathbb{X})\circ\mathbb{Y}\longrightarrow\mathbb{R}$ is the loss function for evaluating the classification performance.

\noindent\textbf{Poisoning-based backdoor attack.} To implement poisoning-based backdoor attacks in supervised learning models, the attacker first designs a trigger injection function $\mathcal{T(\mathbf{X})=\hat{\mathbf{X}}}$, which injects the trigger into the original image $\mathbf{X}$ and generates a poisoned image $\hat{\mathbf{X}}$. Subsequently, the attacker contaminates a subset of the training dataset $\mathbb{X}_{\rm{sub}}\subset\mathbb{X}$ and adjusts the associated labels of selected images to the target label. Following the training of the target model on the tainted dataset, denoted as $\mathcal{F}'$, its behavior undergoes modification, resulting in: $\mathcal{F}'(\mathbf{X})=\rm{y}$, $\mathcal{F}'(\mathcal{T}(\mathbf{X}))=\rm{y}_{\rm{tar}}$.

\noindent\textbf{Clean-image backdoor attack.} Building on the original definition of a backdoor attack, the clean image attack further restricts triggers from being detected by humans. This can be achieved in two ways: (1) by selecting normal semantic objects as triggers, and (2) by limiting the size of the trigger perturbation. The attacker then injects imperceptible perturbations $\mathbf{P}$ into the poisoned dataset $\mathbb{X}_{\rm{sub}}$ and changes the corresponding labels of the poisoned examples. After training, when fed with specific images $\hat{\mathbf{X}}$, the backdoor model $\mathcal{F}$ generates pre-determined predictions $\rm{y}_{\rm{tar}}$ set by the attacker. For INK-I, we selected the second approach by limiting the perturbation size.

\noindent\textbf{Clean-label backdoor attack.} Clean-label backdoor attacks do not manipulate the label of poisoned examples. Denote the selected trigger as $\mathbf{t}$, the attacker selects a series of images $\{\mathbf{X}_1^c,\mathbf{X}_2^c,...,\mathbf{X}_k^c\}$ from the target class $c$ and add adversarial perturbations $\{\mathbf{p}_1,\mathbf{p}_2,...,\mathbf{p}_k\}$ to these examples to destroy the natural feature structure. Then the attacker injects the trigger into perturbed images so that the victim model will consider the trigger as a salient feature for the target class.

\subsection{Capabilities}

\noindent\textbf{Defenders' capabilities.} Defenders have an unauthenticated dataset with lager scales and a trusted dataset to train the teacher model and to perform model distillation, respectively. Defenders can also utilize image-augmentation-based defenses and trigger-inversion-based defenses to detect or disable backdoors.

\noindent\textbf{Attackers' capabilities.} Attackers have access to the unauthenticated dataset and can poison a small portion of images in the dataset. Attackers have no knowledge about the model architecture and have no access to the training process and model distillation process.

%========================================================================================================================

\section{Inheritable Natural Backdoor Attack}
\label{methodology}

Existing poisoning-based backdoor attacks require a direct access to at least a subset of the training dataset, which can not be satisfied in model distillation scenarios. Additionally, the robustness of backdoor attacks faces the challenge of image augmentation and trigger inversion defenses. To meet these requirements, we introduce INK, An Inheritable Natural Backdoor Attack against Model Distillation. Fig.~\ref{Fig_2} provides an overview of the workflow of INK, which consists of two different implementation strategies for clean image attacks (INK-I) and clean label attacks (INK-L), respectively.

In the rest part of this section, we first introduce statistical-feature-based trigger and theoretically analyze its robustness against image augmentations. Subsequently, we provide detailed implementation of INK-I and INK-L. Building upon the implementation of INK, we further theoretically proved the robustness of INK against model distillation.

\subsection{Statistical-based Natural Trigger}

\subsubsection{Trigger Selection}

As discussed above, backdoor attackers have no access to the trusted dataset in model distillation process. However, poisoning-based backdoor attacks require to embed backdoor knowledge by injecting corresponding triggers to the training dataset. Therefore, a practicable solution is to make sure that the selected trigger has already distributed in the trusted dataset naturally. 

Motivated by the insight, we observe that the statistical characteristics of images can be used as natural triggers. These characteristics encompass values such as brightness, saturation, contrast, etc., which can be calculated from a given image, and can divide the training dataset into different categories according to the statistical value. Specifically, we leverage the image variance as the selected trigger, with the corresponding calculation method outlined in Eq.~\ref{VSF}. For an image $x$ with input size $(c, h, w)$, where $c$, $h$, and $w$ represent the number of color channels, height, and width of the image, respectively, this equation calculates its variance $S(x)$ and then multiplies it by an amplification factor $a$.

\begin{equation}
\begin{aligned}
\label{VSF}
    &V_{SF}(x)=a S(x) \\
    &S(x)=\frac{1}{h  w-1}\sum\limits_{j=1}^{h}\sum\limits_{k=1}^{w}(G_{jk}(x)-E(x))^{2} \\
    &E(x)=\frac{1}{h  w}\sum\limits_{j=1}^{h}\sum\limits_{k=1}^{w}G_{jk}(x) \\
    &G_{jk}(x)=\frac{1}{c}\sum\limits_{i=1}^{c} P_{ijk}(x)
\end{aligned}
\end{equation}

\noindent where $P_{ijk}(x)$ represents the pixel value in the $i$-th channel, $j$-th row, and $k$-th column of the image $x$. $G_{jk}(x)$ is the pixel value in the grayscale image. $E(x)$ is the mean value of the grayscale image. The final output $V_{SF}(x)$ represents the image variance value of $x$.

\subsubsection{Augmentation Robustness}

Triggers are robust against image transformations and augmentations if it remains stable under transformations. Formally, for a given transformation denoted as $\mathcal{T}(\cdot)$, INK is robust to $\mathcal{T}(\cdot)$ if the image variance value $V_{SF}(x)$ of image $x$ remains stable after transformation, i.e., $V_{SF}(\mathcal{T}(x))\approx V_{SF}(x)$. Here are theoretical analysis for the robustness of INK against different image augmentations:

\noindent\textbf{Flipping and clipping.} Flipping and clipping will change the spatial structure of images and can deactivate triggers with certain spatial structures. However, $V_{SF}(x)$ will remain unchanged after flipping and clipping because both transformations have no influence on the distribution of images.

\noindent\textbf{Gaussian noises.} Adding Gaussian noises to images will introduce a shift on the distribution. However, this shift has less influence on the image variance because the size of noises is constrained to keep the image quality. Approximating the original distribution of the image by a set of sums of $n$ Gaussian distributions, then the variance can be represented as $\sum_{i=1}^n\sigma_i^2$, adding Gaussian noise to pixels will change the variance to $\sum_{i=1}^n\sigma_i^2 + \alpha^2\sigma^2$, where the weight parameter $\alpha$ is set to be a small value (e.g., 0.01) to keep the image quality. Therefore, we have $\sum_{i=1}^n\sigma_i^2 + \alpha^2\sigma^2 \approx \sum_{i=1}^n\sigma_i^2$. According to Eq.~\ref{VSF}, $V_{SF}(x)$ is positively correlated with the variance, so adding Gaussian noises have limited influence on INK.

\noindent\textbf{Rotation and distortion.} Rotation or distortion will change the distribution of the pixels, for example, for a $45^{\circ}$ rotation transformation, the pixel values in the image's corners will become 0.

We abstract the effect of such transformations as the problem of missing samples in the sample set.
Specifically, for a sample set $x=\{g_{i}|i=1,2,...,n\}$ of grayscale pixels with sample capacity $n=h\times w$, the members in this sample set obey distribution $G\sim N(\mu_{g},\sigma_{g}^{2})$. After adding the transformation $\mathcal{T}(\cdot)$ to $x$, the $r$ percent of pixels is replaced with pixel 0, i.e., $\mathcal{T}(x)=\{g_{ti}|i=1,2,...,n \}=\{g_{i}|i=1,2,... ,(1-r)n\}+\{0_{j}|j=1,2,... ,rn\}$.

In the above case, for the image variance values $V_{SF}(\cdot)$ of $x$ and $\mathcal{T}(x)$, we have the following relationship:
\begin{equation}
    \begin{aligned}
    \label{eq_rotation}
    V_{SF}(\mathcal{T}(x)) \geq (1-r)(V_{SF}(x)-a)
    \end{aligned}
\end{equation}
where $a$ is the amplification factor, as shown in Eq.~\ref{VSF}. 
The proof of Eq.~\ref{eq_rotation} is as follows:
\begin{equation}
    \begin{aligned}
    V_{SF}(\mathcal{T}(x))&=aS(\mathcal{T}(x)) \\
    &=a\times \frac{1}{n-1}\sum\limits_{i=1}^{n}(g_{ti}-m_{t})^2 \\
    &=\frac{a}{n-1}\sum\limits_{i=1}^{(1-r)n}(g_{i}-m_{t})^2 + \frac{a}{n-1}\sum\limits_{j=1}^{rn}(0_{j}-m_{t})^2 \\
    &\geq \frac{a}{n-1}\sum\limits_{i=1}^{(1-r)n}(g_{i}-m_{t})^2 \\
    \end{aligned}
\end{equation}
Since:
\begin{equation}
    \begin{aligned}
    m_{t}&=\frac{1}{n}\sum\limits_{i=1}^{n}g_{ti} =\frac{1}{n}\sum\limits_{i=1}^{(1-r)n}g_{i}+\frac{1}{n}\sum\limits_{j=1}^{rn}0_{j} \\
    &=\frac{1-r}{(1-r)n}\sum\limits_{i=1}^{(1-r)n}g_{i} =(1-r)m
    \end{aligned}
\end{equation}
We have:
\begin{equation}
    \begin{aligned}
    V_{SF}(\mathcal{T}(x)) &\geq \frac{a}{n-1}\sum\limits_{i=1}^{(1-r)n}(g_{i}-m_{t})^2 \\
    &= \frac{a}{n-1}\sum\limits_{i=1}^{(1-r)n}(g_{i}-(1-r)m)^2 \\
    &= \frac{a}{n-1}\sum\limits_{i=1}^{(1-r)n}(g_{i}-m)^2+\frac{a}{n-1}\sum\limits_{i=1}^{(1-r)n}2rg_{i}m  \\
    &\quad + \frac{a}{n-1}\sum\limits_{i=1}^{(1-r)n}(1-r)^{2} - \frac{a}{n-1}\sum\limits_{i=1}^{(1-r)n}1\\
    &\geq \frac{a}{n-1}\sum\limits_{i=1}^{(1-r)n}(g_{i}-m)^2 - \frac{(1-r)an}{n-1} \\
    &\approx (1-r)(V_{SF}(x)-a)
    \end{aligned}
\end{equation}

Once we determined the effect of $\mathcal{T}(\cdot)$ on $V_{SF}(x)$, the attacker can use redundant $V_{SF}(x)$ to remove the impact of  $\mathcal{T}(\cdot)$.
Specifically, for an image $x$, the attacker uses Eq.~\ref{Activation} to increase the $V_{SF}(x)$ of $x$ to make:
\begin{equation}
    \begin{aligned}
    &V_{SF}(\mathcal{T}(x_{t}))\geq th
    \end{aligned}
\end{equation}

i.e.,

\begin{equation}
    \begin{aligned}
    V_{SF}(\mathcal{T}(x_{t}))&\geq (1-r)(V_{SF}(x_{t})-a) \\
    &=(1-r)(V_{SF}(F(x,\gamma,\lambda))-a) \geq th
    \end{aligned}
\end{equation}

Combining the above theoretical analysis, we can conclude that INK is robust against image augmentations.

\subsection{Clean Image Attack (INK-I)}

In clean image attack scenarios, attackers can add unrecognizable triggers to a subset of the training dataset, and flip the corresponding labels of poisoned examples. The implementation of INK under this scenarios is named as INK-I. As illustrated in Fig.\ref{Fig_2}, the data poisoning process of INK-I can be detailed as follows:

(1) Initially, the attacker utilizes Eq.~\ref{VSF} to calculate the image variance values $V_{SF}(x)$ for images in the training dataset with the input size of $(c, h, w)$.

(2) Next, the attacker sets the image variance threshold $th$ according to the targeted poisoning ratio $r_{p}$. (e.g., if the poisoning ratio $r_{p}$ is set to $5\%$, then the threshold should be set to the value that ranks in the $5\%$ or $95\%$ of the image variance values of the training samples).

(3) If the image variance value $V_{SF}(x)$ of a training image surpassing the threshold $th$, the attacker flips its corresponding label to the target label $\eta(y)$.

Algorithm\ref{alg_clean_image} presents the pseudocode for the poisoning process of INK-I.

After training on the poisoned dataset, the model will build up a malicious connection between the image variance and the target label. An attacker can exploit this correlation through a specific trigger and control the behavior of the victim model.

\begin{figure*}[htbp]
\centering
\includegraphics[width=0.9\linewidth]{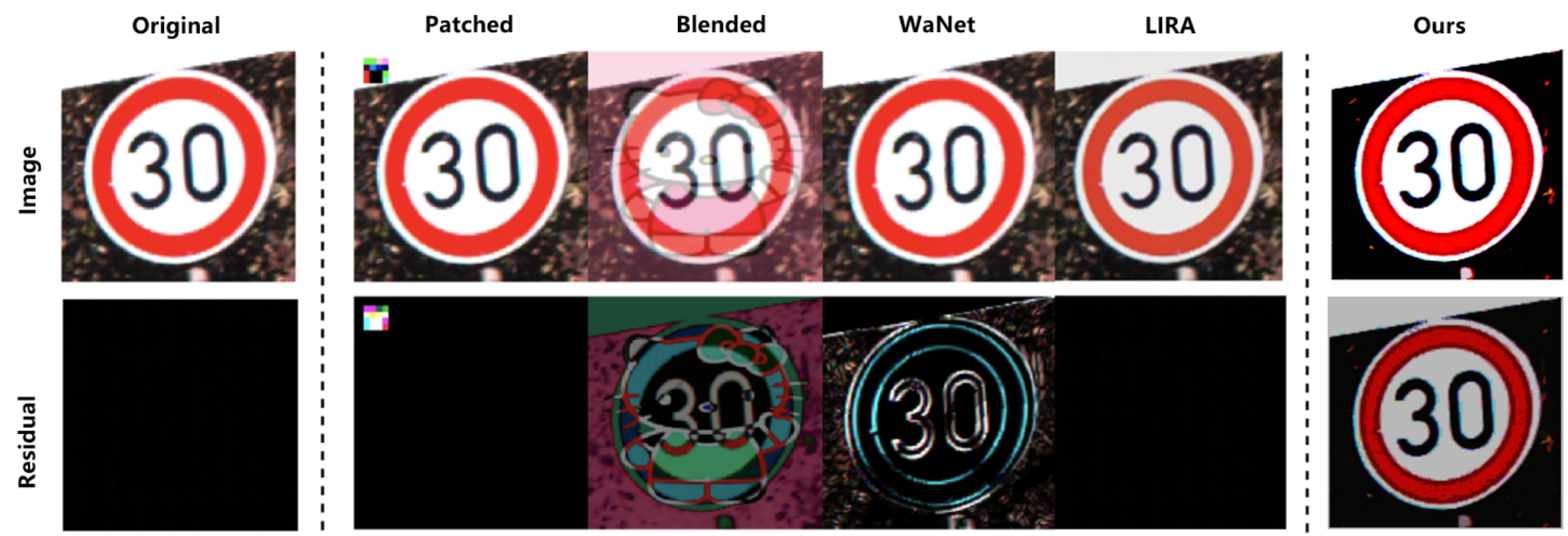}%
\caption{Visualization of backdoor activation, including images generated by patch-based BadNets~\cite{ChenTB17}, Blended~\cite{LiuM0020}, WaNet~\cite{NguyenWN21}, LIRA~\cite{DoanLI21}, and our INK.}
\label{triggers1}
\end{figure*}

\begin{algorithm}[htbp]
\caption{INK-I}\label{alg_clean_image}
\KwIn{Target Dataset $D_{tar}$, Label Converter $\eta(\cdot)$, Poisoning Ratio $r_{p}$}
\KwOut{Poisoned dataset $D_{poi}$}
Initialize $D_{tar}$ = $empty(\ )$\;
Determining the threshold $th$ based on $r_{p}$ \;
\For{$(x,y)$ = $iter(D_{tar}).next(\ )$}
{
    \eIf{$V_{SF}(x) \geq th$}
        { 
            $D_{poi}.append((x,\eta(y)))$\;
        }
        {
            $D_{poi}.append((x,y))$\;
        }
}
\end{algorithm}

\noindent\textbf{Backdoor activation.} As described in the above description of backdoor injection, backdoors established by INK-I are associated with image variance values. During an attack, the attacker can activate the backdoor by enhancing the \(V_{SF}(x)\) of the image above \(th\).

In this problem, we enhance the image variance value of the image \(x\) by using the transformation function \(F\). As shown in Eq.~\ref{Activation}, this function enhances the contrast between larger pixels and smaller pixels, thus increasing the image variance value:

\begin{equation}
\begin{aligned}
\label{Activation}
&x_t=F(x,\gamma,\lambda) \\
&\mathrm{ s.t. }\  V_{SF}(x_{t})\geq th \\
&\qquad P_{ijk}(x_{t})=\lambda \cdot (P_{ijk}(x_{r}))^{\gamma} + (1-\lambda) \cdot P_{ijk}(x) \\
&\qquad P_{ijk}(x_{r})=P_{ijk}(x)-\min\{P_{i}(x)\}
\end{aligned}
\end{equation}

\noindent where (\(\gamma\) is the trade-off parameter for the equilibrium transformation. This change amplifies \(V_{SF}(x)\) when \(\gamma > 1\) and \(\lambda\) denotes the mixing ratio used to enhance visual concealment. Fig.~\ref{triggers1} provides some visualized examples of triggered images generated using Eq.~\ref{Activation}.

Based on Eq.~\ref{Activation}, attackers can further design customized activation methods such as adding sparse perturbations. It is worth noting that the trigger generated by INK is sample-specific and natural, which enables it to bypass trigger-inversion-based backdoor detection algorithms. 

\begin{algorithm}[htbp]
\caption{INK-L}\label{alg_clean_label}
\KwIn{Target Dataset $D_{tar}$, Label Converter $\eta(\cdot)$,
Threshold $th$,
Poisoned Number $n$, 
Benign Model $\mathcal{C}$}
\KwOut{Poisoned Dataset $D_{poi}$}
Initialize $D_{poi}$ = $empty(\ )$\;
\For{$(x,y)$ = $iter(D_{tar}).next(\ )$}
{
    \eIf{$y \neq \eta(y)$}
        { 
            $x_{temp}$ = $F(x,\gamma_{1},\lambda_{1})$\;
            \eIf{$V_{SF}(x)$ $\geq$ $th$}{
                $D_{poi}.append((x_{temp},y))$\;
            }
            {
                $D_{poi}.append((x,y))$\;
            }
        }
        {
            $x_{temp}$ = $untarget\_PGD(x,y,\mathcal{C})$\;
            $x_{temp}$ = $F(x_{temp},\gamma_{2},\lambda_{2})$\;
            \eIf{
                $n > 0$
                $\&$ $V_{SF}(x_{temp}) \geq th$ 
                $\&$ $\mathcal{C}(x_{temp}) \neq \eta(y)$
            }{
                $n$ = $n$ - 1\;
                $D_{poi}.append((x_{temp},y))$\;
            }
            {
                $D_{poi}.append((x,y))$\;
            }
        }
}
\end{algorithm}

\begin{figure*}[tbp]
\centering
\includegraphics[width=0.9\linewidth]{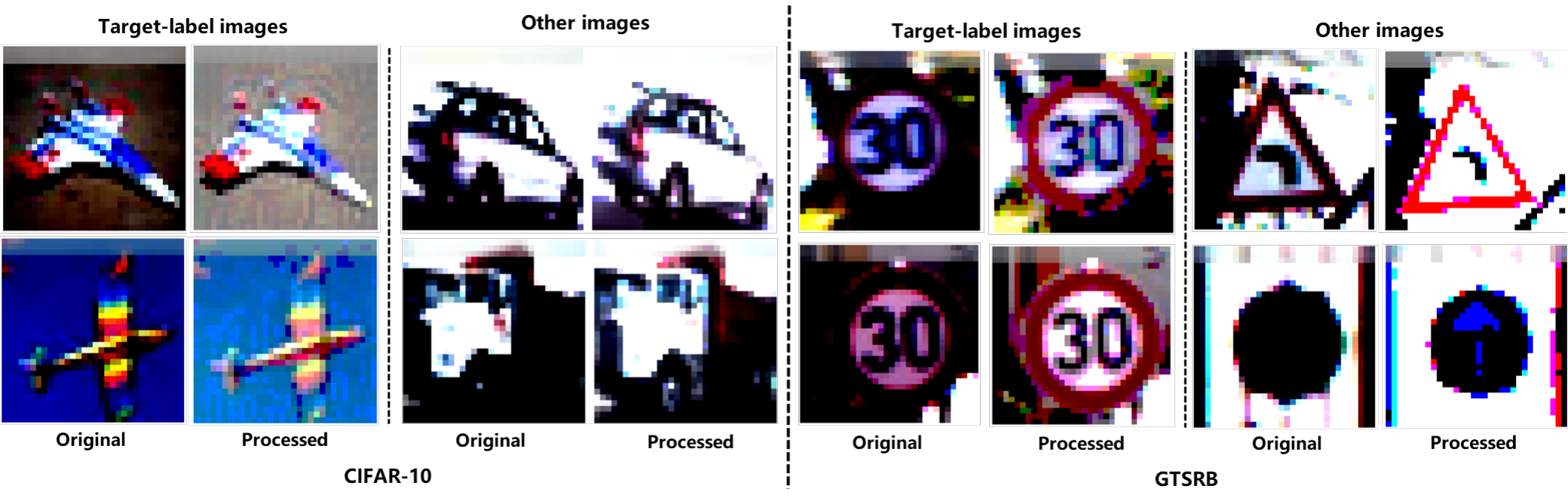}
\caption{Visualization of INK-L implantation. Columns $1\sim4$ are the images from CIFAR-10, and Columns $5\sim 8$ are the images from GTSRB. we use Algorithm~\ref{alg_clean_label} to process the datasets. Algorithm~\ref{alg_clean_label} filters and processes specific images in the training sets. Images before and after processing are shown above. Visually, the processed images do not look unusual.}
\label{Clean_label}
\end{figure*}

\subsection{Clean Label Attack (INK-L)}

\subsubsection{Backdoor Injection}

In clean label attack scenarios, attackers can add triggers to a subset of the training dataset, but can not flip the corresponding labels of poisoned examples. The implementation of INK under this scenarios is named as INK-L. As illustrated in Fig.\ref{Fig_2}, the data poisoning process of INK-L can be detailed as follows:

(1) Initially, the attacker determines an upper bound $n$ on the number of labeled images of the poisoning target and a threshold $th$ for $V_{SF}(x)$.

(2) For $x$ in the training set,
\begin{itemize}
    \item If the corresponding label $y$ stisfies $y\neq\eta(y)$ and $V_{SF}(x)$ exceeds the threshold $th$, INK-L utilizes $F(x,\gamma_{1},\lambda_{1})$ to increase its image variance value below the threshold $th$ by setting $\gamma_{1}$.
    
    \item If the corresponding label $y$ stisfies $y=\eta(y)$ and the count of currently poisoned target-label images is below $n$, INK-L execute an untargeted PGD~\cite{MadryMSTV17} attack on $x$ using a benign local model. The optimization problem is defined in Eq.~\ref{PGD_cl}, where the perturbation $\delta$ aims to disrupt $x$'s prediction from aligning with the ground truth $y$, and $\epsilon$ constrains the magnitude of $\delta$.
    \begin{equation}
    \label{PGD_cl}
        \begin{aligned}
        \max\limits_{\delta \in \Delta} Loss(\mathcal{C}(x+\delta),y), \ 
        \Delta = \{\delta \in \mathbb{R}^{c,h,w},||\delta||_{p} \leq \epsilon \}
        \end{aligned}
    \end{equation}
    
    \item Subsequently, INK-L increase the image variance value of $x$ beyond the poisoning threshold $th$ using $F(x,\gamma_{2},\lambda_{2})$.

\end{itemize}

Algorithm\ref{alg_clean_label} presents the pseudocode for the poisoning process of INK-L. As dipicted in Fig.~\ref{Clean_label}, the content of poisoned images generated by INK-L align with their labels.

\noindent\textbf{Backdoor Activation.} Similar to INK-I, the  attacker activates the backdoor injected by INK-L by using Eq.~\ref{Activation} to increase $V_{SF}(x)$

%========================================================================================================================

\subsection{Distillation Robustness}

During the model distillation process, the backdoor attacker lacks control over the distillation process or access to the trusted dataset, leading to the failure of existing poisoning-based backdoor attacks. 

As for our INK, it is robust against model distillation because the $V_{SF}(x)$ distribution is similar across different datasets. This means that the poisoned image exists in all these datasets, which makes the transfer of INK during model distillation achievable. The detailed proof is as follows:

 For an image $x$ with size $(c,h,w)$, denoting the distribution of its pixel value $p$ as $P\sim \mathcal{N}(\mu, \sigma ^{2})$, then, the distribution of gray scale pixel value $p_{g}$ in  corresponding grayscale image $x_{g}$ could be denoted as $G\sim \mathcal{N}(\mu_{g},\sigma_{g}^{2})$, where $\mu_{g} = \mu, \sigma_{g}^{2}=\frac{\sigma^{2}}{c}$. According to the Central Limit Theorems, the variance $S(x)$ satisfies:
\begin{equation}
    \begin{aligned}
    \frac{(n-1)  S(x)}{\sigma_{g}^{2}} \sim \mathcal{X}^{2}(n-1) \quad ,\quad n=h\times w
    \end{aligned}
\end{equation}
Therefore, we have:
\begin{equation}
    \begin{aligned}
    V_{SF}(x)=a  S(x) \sim \frac{a  \sigma_{g}^{2}}{n-1} \mathcal{X}^{2}(n-1)
    \end{aligned}
\end{equation}
Taking in the variables, we have:
\begin{equation}
    \begin{aligned}
    V_{SF}(x) \sim \frac{a  \sigma^{2}}{c (h w-1)} \mathcal{X}^{2}(h  w - 1 )
    \end{aligned}
\end{equation}

It can be seen that the distribution of $V_{SF}(x)$ is related to the standard deviation $\sigma$ and the image size $c,h,w$. For a model, the size of its input image is generally fixed, so we only need to consider the influence of $\sigma$. According to our assumptions in the threat model, the victim will standardize the training set when training the model, i.e., $\sigma = 1$. Therefore, the distributions of $V_{SF}(x)$ are approximately the same for different data sets, so our backdoor can transfer across data sets, i.e., INK is robust to model distillation.

%========================================================================================================================

\section{Experiments}
\label{experiments}

\subsection{Experimental Setup}

\noindent\textbf{Dataset.} Building upon previous studies on backdoor attacks, we opted for two extensively utilized datasets: CIFAR-10~\cite{krizhevsky2009learning} and GTSRB~\cite{HoubenIJCNN2013}. CIFAR-10 comprises a diverse collection of 60,000 32x32 color images, evenly distributed across 10 distinct categories which encompass common objects and animals. The German Traffic Sign Recognition Benchmark (GTSRB) focuses on traffic sign classification. This dataset includes over 50,000 images of traffic signs, spanning 43 different classes. GTSRB is widely recognized for its relevance to real-world applications, particularly in the development of systems for autonomous vehicles and traffic management.

In the distillation experiment, we introduced the CINIC-10 dataset~\cite{Darlow18CINIC} to evaluate the robustness of INK to model distillation. CINIC-10 is a composite dataset, combining images from CIFAR-10, ImageNet, and other sources, resulting in a diverse collection of 270,000 images distributed across 10 classes.

\noindent\textbf{Model.} For experimental validation, we primarily utilize the popular Pre-activation ResNet-18 model~\cite{HeZRS16}. Additionally, our comparative experiments include PreActResNet34 and PreActResNet50, offering deeper networks with 34 and 50 layers, respectively. The pre-activation design enhances gradient flow, contributing to more effective training of deep neural networks. VGG16~\cite{SimonyanZ14a}, composed of 16 weight layers, is known for its simplicity and effectiveness in image classification tasks. GoogLeNet~\cite{SzegedyLJSRAEVR15}, also known as Inception-v1, introduces inception modules, enabling the network to capture features at different spatial scales.

\noindent\textbf{Baselines.} Since INK have two implementation strategies for both clean image and clean label attacks, we select two clean image backdoor attacks (WaNet~\cite{NguyenWN21} and LIRA~\cite{DoanLI21}) and two clean label backdoor attacks (SAA~\cite{SouriFCGG22} and NARCISSUS~\cite{ZengPJLQJ23}) as baselines. WaNet~\cite{NguyenWN21} employs a warp function to generate triggers that conform to the contours of images, making them challenging for humans to detect. 
LIRA~\cite{DoanLI21} attains significant concealment by framing the trigger generation process as an optimization problem. SAA~\cite{SouriFCGG22} employs gradient matching, data selection, and target model re-training to backdoor models trained from scratch. NARCISSUS~\cite{ZengPJLQJ23} utilizes out-of-distribution data to enhance the effectiveness of clean label attacks.

All baselines achieve human-imperceptibility and high attack success rates. Meanwhile, they are proposed to be robust against existing backdoor detection methods. We follow the official implementation details to reproduce these baseline methods and compare them with INK in different scenarios. 

\noindent\textbf{Metrics.} We assess the performance of INK and baseline methods under various attack settings using two widely employed metrics: Benign Accuracy (\textbf{ACC}) and Attack Success Rate (\textbf{ASR}).

\noindent\textbf{Hyperparameters.} The hyperparameters associated with backdoor implantation and activation were determined through trial and error. In Eq.~\ref{VSF}, the amplification factor $a$ is set to 100. We employ SGD optimizers with a learning rate of 0.01 for network training. In the case of INK-I, unless explicitly mentioned, we utilize a poisoning ratio $r_{p}$ of 1\% across all experiments. The hyperparameters for INK-L, as outlined in Algorithm\ref{alg_clean_label}, are detailed in Tab.~\ref{cl_para}. In the case of untargeted PGD in Eq.~\ref{PGD_cl}, we execute 15 rounds of updation on the target-label images, with a step size of 0.01.

\begin{table}[htbp]
\caption{Hyperparameters of INK-L.}
\label{cl_para}
\centering
\begin{tabular}{ccc}
\hline
\multirow{2}{*}{\makebox[0.15\textwidth]{Parameter}}             & \multicolumn{2}{c}{Value}            \\ \cline{2-3} 
                                       & \makebox[0.1\textwidth]{CIFAR-10} & \multicolumn{1}{l}{\makebox[0.1\textwidth]{GTSRB}} \\ \hline
target-label                           & 0        & 1                         \\
$n$                                    & 250      & 200                       \\
$th$                                   & \multicolumn{2}{c}{160}              \\
$\mathcal{C}$                          & \multicolumn{2}{c}{PreRes18}         \\
$\gamma_{1}$                           & \multicolumn{2}{c}{0.7}              \\
$\lambda_{1}$                          & \multicolumn{2}{c}{1}                \\
$\gamma_{2}$                           & \multicolumn{2}{c}{2.5}              \\
$\lambda_{2}$                          & \multicolumn{2}{c}{0.5}              \\ \hline
\end{tabular}
\end{table}

\subsection{Attack Perfomance}

\noindent\textbf{Overall comparison.} Initially, we compare the attack performance of INK with baseline attacks under no-defense scenarios. The evaluation results are listed in Tab.~\ref{Tab_2}.

For clean image attacks, INK-I achieves a comparable attack success rate (0.979 on average) with WaNet and LIRA (0.975 and 0.978 on average, respectively), while maintaining a high usability on benign inputs. Although LIRA has a slightly higher ACC and ASR, it requires attackers to have control over the training process of the victim model, which makes it less practicable in real-world scenarios.

\begin{table}[htbp]
\caption{Performance comparison without defense.}
\label{Tab_2}
\resizebox{\linewidth}{!}{%
\begin{tabular}{@{}llcccccc@{}}
\toprule[2pt]
\multirow{2}{*}{Dataset}  & \multirow{2}{*}{Model} & \multicolumn{2}{c}{WaNet} & \multicolumn{2}{c}{LIRA}      & \multicolumn{2}{c}{INK-I} \\
                          &                        & ACC         & ASR         & ACC            & ASR          & ACC         & ASR         \\ \midrule
\multirow{2}{*}{CIFAR-10} & PreRes-18              & 0.858       & 0.986       & 0.862          & 0.952        & 0.857       & 0.997       \\
                          & VGG-16                 & 0.849       & 0.992       & 0.857          & 0.985        & 0.851       & 0.983       \\ \midrule
\multirow{2}{*}{GTSRB}    & PreRes-18              & 0.964       & 0.959       & 0.962          & 0.987        & 0.960       & 0.962       \\
                          & VGG-16                 & 0.962       & 0.963       & 0.960          & 0.990        & 0.960       & 0.975       \\ \midrule[1.5pt]
\multirow{2}{*}{Dataset}  & \multirow{2}{*}{Model} & \multicolumn{2}{c}{SAA}   & \multicolumn{2}{c}{NARCISSUS} & \multicolumn{2}{c}{INK-L} \\
                          &                        & ACC         & ASR         & ACC            & ASR          & ACC         & ASR         \\ \midrule
\multirow{2}{*}{CIFAR-10} & PreRes-18              & 0.852       & 0.786       & 0.854          & 0.990        & 0.856       & 0.989       \\
                          & VGG-16                 & 0.855       & 0.842       & 0.858          & 0.981        & 0.852       & 0.988       \\ \midrule
\multirow{2}{*}{GTSRB}    & PreRes-18              & 0.959       & 0.903       & 0.965          & 0.991        & 0.963       & 0.985       \\
                          & VGG-16                 & 0.963       & 0.885       & 0.953          & 0.999        & 0.960       & 1.000       \\ \bottomrule
\end{tabular}}
\end{table}

As for clean label attacks, INK-L achieves an attack success rate of 0.990 on average, which outperforms SAA (0.854 on average) and is close to NARCISSUS (0.990 on average).  

\begin{figure}
\centering
\includegraphics[width=\linewidth]{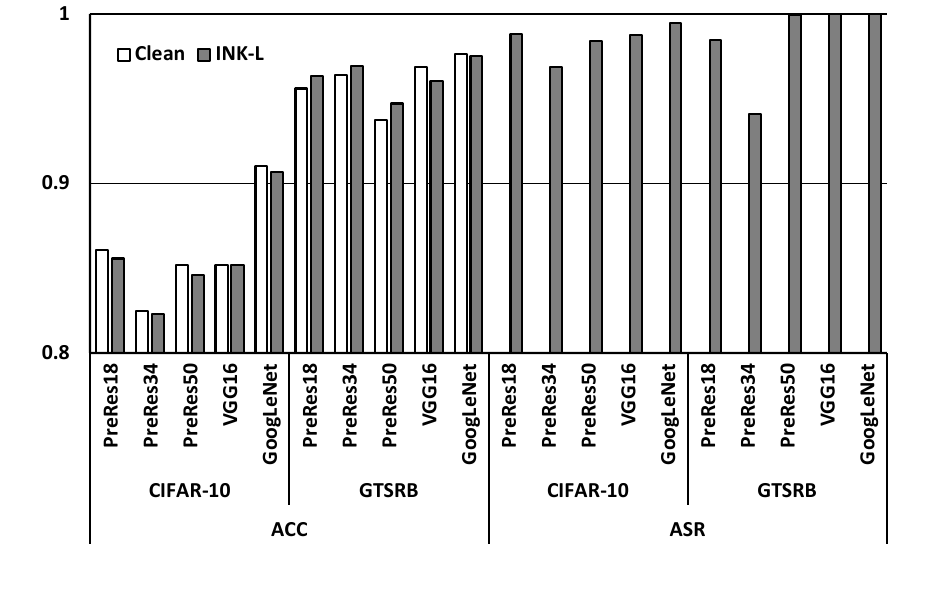}
\caption{Extended evaluation of INK-L on different datasets and models.}
\label{cl_blackbox}
\end{figure}

\noindent\textbf{Black-Box Clean-Label Attack.} Most previous clean label attacks require prior knowledge about the model architecture, or the access to the training process of the victim model. INK-L does not rely on these prior knowledge and can be applied under black-box constraints. Therefore, we conducted additional experiments to evaluate the performance of INK-L on different datasets and models. 

Specifically, INK-L initially utilizes PreActResNet18 (model $\mathcal{C}$ in Algorithm \ref{alg_clean_label}) to generate a poisoned clean label dataset. This dataset was then employed to train backdoor models (model $\mathcal{B}$) with different architecture including PreActResNet18, PreActResNet50, VGG16, and GoogLeNet. As depicted in Fig.~\ref{cl_blackbox}, INK-L achieves an ASR exceeding 95\% on average without knowing the specific model architecture or having control over the training process.  

The overall performance comparison of INK-I and INK-L with baseline methods demonstrates the effectiveness of INK under no-defense scenarios. Subsequently, we will evaluate the robustness of INK under different defenses.

\subsection{Augmentation Robustness}

\begin{figure}[htbp]
\centering
\includegraphics[width=0.6\linewidth]{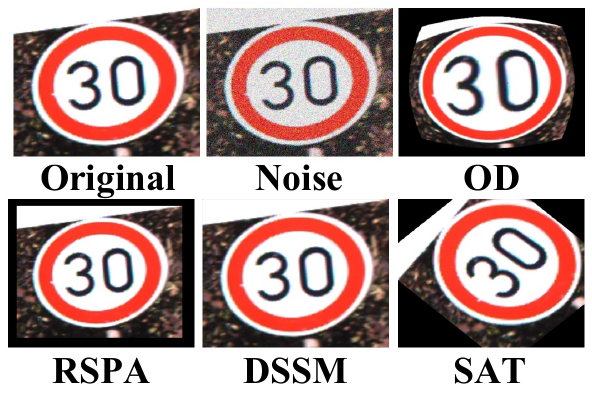}
\caption{Visualized examples of the image augmentations used for evaluation.}
\label{Augmentation_image}
\end{figure}

We theoretically analyzed the robustness of INK against image augmentations in Sec.~\ref{methodology}. Additionally, we compare the attack performance of INK with selected baselines under different image augmentation algorithms in this section. 

Specifically, we use four image augmentation operations, Optical Distortion (OD), Stochastic Affine Transformation (SAT), Random scaling down with Padding (RSPA), and Median filter with Scaling Down (DSSM), to assess the robustness of backdoor attacks against image augmentations. Additionally, we explored the impact of Gaussian noise on triggers. The image augmentation techniques used in the experiment are showcased in Fig.~\ref{Augmentation_image}.

Tab.~\ref{Augmentation_table} lists the attack success rates of different backdoor attacks under different image augmentations.
As shown in Tab.~\ref{Augmentation_table}, INK-I and INK-L maintain a high attack success rate after different image augmentations, which supports the theoretical analysis. Additionally, compared to WaNet and SAA, LIRA and NARCISSUS are overally more robust against image augmentations. This is because LIRA and NARCISSUS require backdoor attackers to control to training process and already introduced several image augmentations during training. However, augmentation strategies that have not been introduced during training may have a significant impact on the attack success rate (e.g., adding Gaussian noise for LIRA).

\begin{table}[tbp]
\caption{Attack success rates under different image augmentations.}
\centering
\resizebox{\linewidth}{!}{%
\begin{tabular}{@{}cccccccc@{}}
\toprule
Dataset                   & Augmentation & WaNet & LIRA  & SAA   & NARCISSUS & INK-I & INK-L \\ \midrule
\multirow{6}{*}{CIFAR-10} & None         & 0.986 & 0.952 & 0.786 & 0.990     & 0.997 & 0.989 \\
                          & Noise        & 0.589 & 0.135 & 0.523 & 0.938     & 0.993 & 0.967 \\
                          & OD           & 0.624 & 0.987 & 0.469 & 0.943     & 0.977 & 0.954 \\
                          & SAT          & 0.395 & 0.961 & 0.775 & 0.902     & 0.989 & 0.981 \\
                          & RSPA         & 0.519 & 0.961 & 0.452 & 0.945     & 0.990 & 0.982 \\
                          & DSSM         & 0.399 & 0.984 & 0.375 & 0.952     & 0.968 & 0.937 \\ \midrule
\multirow{5}{*}{GTSRB}    & None         & 0.959 & 0.987 & 0.903 & 0.991     & 0.962 & 0.985 \\
                          & Noise        & 0.03  & 0.854 & 0.797 & 0.858     & 0.982 & 0.898 \\
                          & OD           & 0.616 & 0.993 & 0.676 & 0.874     & 0.869 & 0.758 \\
                          & SAT          & 0.71  & 0.933 & 0.855 & 0.738     & 0.953 & 0.868 \\
                          & RSPA         & 0.478 & 0.996 & 0.558 & 0.879     & 0.972 & 0.900 \\
                          & DDSM         & 0.199 & 0.857 & 0.463 & 0.767     & 0.876 & 0.783 \\ \bottomrule
\end{tabular}}
\label{Augmentation_table}
\end{table}

\subsection{Distillation Robustness}
\label{distill_section}

We then evaluate the robustness of INK against model distillation and compare it with baseline methods. Specifically, we consider four different distillation strategies:

\begin{itemize}
    \item \textbf{S1: Standard.} The teacher model and the student model have the same architecture and are trained on the same dataset;
    \item \textbf{S2: Different Model.} The teacher model and the student model are trained on the same dataset but have different architectures;
    \item \textbf{S3: Different Dataset.} The teacher model and the student model have the same architecture but are trained on different datasets;
    \item \textbf{S4: Different Task.} The teacher model and the student model have the same architecture and trained on similar but not the same dataset.
\end{itemize}

Detailed settings are outlined in Tab.~\ref{distill_secnario}:

\begin{table}[htbp]
\caption{Detailed settings for distillation scenarios.}
\centering
\resizebox{\linewidth}{!}{%
\begin{tabular}{clclccclc}
\Xhline{1px}
Scenarios &  & Teacher &  & Dataset         &  & Student &  & Dataset                \\ \hline
S1        &  & P-Res18    &  & CIFAR-10(GTSRB) &  & P-Res18    &  & CIFAR-10(GTSRB)        \\
S2        &  & P-Res18    &  & CIFAR-10(GTSRB) &  & GoogLeNet    &  & CIFAR-10(GTSRB)        \\
S3        &  & P-Res18    &  & CIFAR-10(GTSRB) &  & P-Res18    &  & CINIC-10 \\
S4        &  & P-Res18    &  & CIFAR-10(GTSRB) &  & P-Res18    &  & CIFAR-9(GTSRB) \\ \Xhline{1px}
\end{tabular}}
\label{distill_secnario}
\end{table}

Tab.~\ref{distill_result} provides the evaluation results of different backdoor attacks. According to the table, all four baseline backdoor attacks fails with high probabilities after different model distillation strategies. Despite the different implementation details, these methods share a fundamental pre-request that attackers must have the access to the training dataset, which is not satisfied in model distillation scenarios. As the result, the transmission of backdoor knowledge from the teacher model to the student model is blocked.

For INK, as analyzed in Sec.~\ref{methodology}, the corresponding triggers have already distributed in the trusted dataset. Therefore, once the attacker successfully backdoored the teacher model, it will become a poisoned labeler during the distillation process, and the model distillation process will change to a label-based backdoor attack process. as shown in Tab.~\ref{distill_result}, after model distillation, INK-I and INK-L have a similar (even higher) attack success rate on the trusted student model compared to the teacher model, which supports the theoretical results.

\begin{table}[htbp]
\caption{Attack success rate comparison after model distillation.}
\centering
\resizebox{\linewidth}{!}{%
\begin{tabular}{@{}ccccccc@{}}
\toprule
Dataset                  & Attack    & Teacher  & S1    & S2    & S3    & S4    \\ \midrule
\multirow{6}{*}{CIFAR10} & WaNet     & 0.986    & 0.029   & 0.017  & 0.021  & 0.001  \\
                         & LIRA      & 0.952    & 0.014   & 0.006  & 0.014  & 0.007  \\
                         & SAA       & 0.786    & 0.017   & 0.029  & 0.035  & 0.018  \\
                         & NARCISSUS & 0.990    & 0.021   & 0.009  & 0.014  & 0.016  \\
                         & RSBA-CI   & 0.997    & 0.995   & 0.998  & 0.991  & 0.990 \\
                         & RSBA-CL   & 0.989    & 0.943   & 0.951  & 0.988  & 0.971 \\ \midrule
\multirow{6}{*}{GTSRB}   & WaNet     & 0.959    & 0.010   & 0.012  & -      & 0.008  \\
                         & LIRA      & 0.987    & 0.026   & 0.004  & -      & 0.013  \\
                         & SAA       & 0.903    & 0.011   & 0.014  & -      & 0.017  \\
                         & NARCISSUS & 0.991    & 0.020   & 0.025  & -      & 0.013  \\
                         & RSBA-CI   & 0.962    & 0.853   & 1.000  & -      & 0.999 \\
                         & RSBA-CL   & 0.985    & 0.960   & 0.952  & -      & 0.944  \\ \bottomrule
\end{tabular}
\label{distill_result}}
\end{table}

\subsection{Backdoor Defenses}

\begin{figure}[htbp]
\centering
\includegraphics[width=0.6\linewidth]{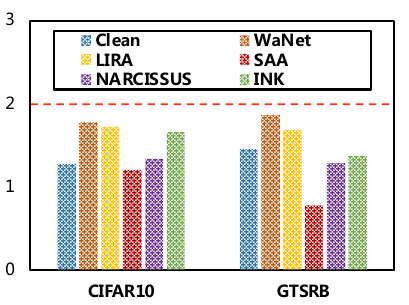}
\caption{Detection results of Neural Cleanse against different backdoor attacks.}
\label{Neural_cleanse}
\end{figure}

\noindent\textbf{Neural Cleanse.} Neural Cleanse searches for the optimal patterns as potential backdoor trigger for each target class. These patterns force the model to classify any clean input to a certain category. Utilizing the Anomaly Index metric, Neural Cleanse assesses whether a label exhibits a significantly smaller pattern, indicating a potential backdoor. Specifically, a backdoor model is identified if its Anomaly Index surpasses the threshold of 2. As depicted by Fig.~\ref{Neural_cleanse}, the Anomaly Index value of INK closely align with those of the clean model.

More importantly, as discussed in Sec.~\ref{methodology}, the trigger generated by INK is sample-specific, which makes searching-based backdoor detection strategies struggle to find a fixed trigger. Therefore, INK is robust against this kind of defenses.

\begin{figure}[htbp]
\centering
\includegraphics[width=\linewidth]{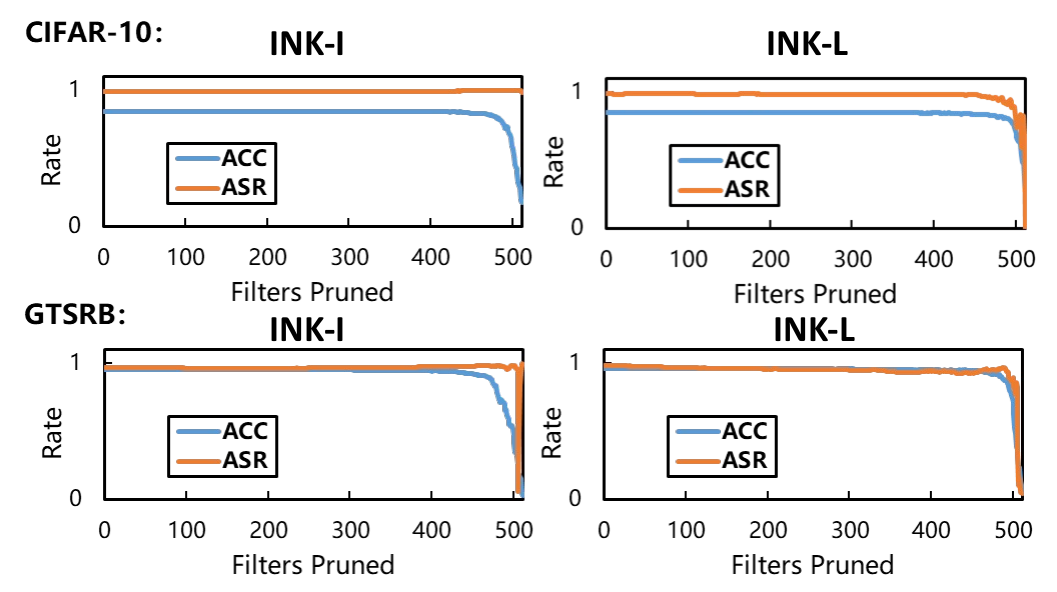}
\caption{Defense performance of Model Fine-pruning.}
\label{Finepruning}
\end{figure}

\noindent\textbf{Model Fine-pruning.} Fine-pruning suggests a connection between the backdoor and specific neurons, eliminating it by gradually pruning those neurons in a designated layer. Experiments on CIFAR-10, as depicted in Fig.~\ref{Finepruning}, show the accuracy (ACC) and attack success rate (ASR) during the pruning process. The consistently high ASR throughout pruning highlights the ability of INK to elude the Fine-pruning defense.

\subsection{INK in Non-Standardization Case.}

To evaluate the scalability of INK, we designed an experiment to verify it in a non-standardization case. Specifically, we use a statistic feature similar to the Coefficient of Variation as the statistical feature in this experiment, as shown in Eq.~\ref{VSF_ns}:  

\begin{equation}
\begin{aligned}
\label{VSF_ns}
    V_{SF}(x)=\frac{a S(x)}{E(x)}
\end{aligned}
\hspace{2em}
\end{equation}

\begin{equation}
\begin{aligned}
\label{activation_ns}
   P_{ijk}(x)=\frac{P_{ijk}(x)-\mu}{\sigma}
\end{aligned}
\end{equation}

\begin{figure}[htbp]
\centering
\includegraphics[width=\linewidth]{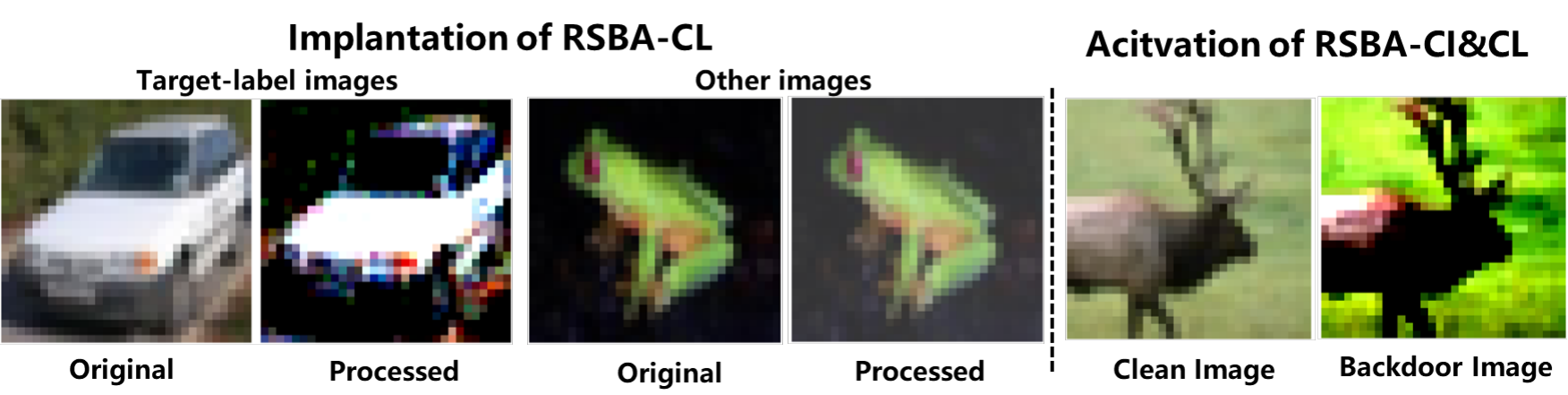}
\caption{Visualized examples of images in non-standardization case.}
\label{std_compare}
\end{figure}

\begin{table}[htbp]
\caption{INK in non-standardization case.}
\centering
\begin{tabular}{cccc}
\Xhline{1px}
\makebox[0.1\textwidth]{DataSet}                   & \makebox[0.1\textwidth]{Method}      & \makebox[0.1\textwidth]{ACC}   & \makebox[0.09\textwidth]{ASR}              \\ \hline
\multirow{3}{*}{CIFAR-10} & Clean Model & 85.73 & - \\
                          & INK-I       & 85.44 & 98.69            \\
                          & INK-L       & 85.41 & 89.54            \\ 
\Xhline{1px}
\end{tabular}
\label{non_s}
\end{table}

For the backdoor activation, we directly use the standardization function as our trigger function since the standardization function will change $E(x)$ to 0, thus increasing $V_{SF}(x)$. The standardization function is shown in Eq.~\ref{activation_ns}, where the $\mu$ and $\sigma$ are the mean value and variance of the training set. Visualization of INK in non-standardization case is shown in Fig.~\ref{std_compare}.

In non-standardization cases, the performance of clean model $\mathcal{C}$ and backdoor model $\mathcal{B}$ is shown in Tab.~\ref{non_s}. It can be observed that without image standardization during the training process, the attacker can also achieve INK implantation and obtain a 98.69\% ASR of INK-I and 89.54\% ASR of INK-L through the design of the trigger.
These results suggests that INK introduces a new perspective for poisoning backdoor attacks, which makes INK more scalable and difficul for defenders to detect.

%========================================================================================================================

\subsection{Discussion and Limitation}

As INK utilizes statistical properties of the training images, it would seem relatively easy to construct an effective defense against the proposed attack. For example, one could filter out any training images that have high values for common statistical properties.

However, the reality is that INK introduces a new attack paradigm rather than being limited to a specific implementation. Attackers don't have to adhere to common statistical values when launching INK. Here are a couple of ways it can be improved:

(1) Frequency domain instead of time domain: The attacker can utilize the frequency domain features of the image instead of the basic time domain features to carry out the attack.

(2) Localized sampling instead of overall sampling: The attacker can compute statistical features by using pixel values from a few specific locations instead of the entire image.

(3) Ensemble strategies: The attacker can use a combination of multiple statistical features as triggers to enhance the robustness and make backdoor detection difficult.

When the defender is unaware of the attacker's specific calculation method, it becomes difficult for the attacker to accurately screen those problematic images.

%========================================================================================================================

\section{Conclusion}
\label{conclusion}

In this paper, we introduce INK, the first statistical-feature-based backdoor attack. INK empowers the attacker to execute a black-box clean-image or clean-label attack without engaging in the model training process. Furthermore, we demonstrate both theoretically and experimentally that our trigger design renders INK resilient to model distillation and image augmentation. Experiments reveal that our method achieves a high attack success rate in black-box scenarios and can elude widely-used backdoor defenses. In future work, we will explore the potential applications of INK in watermarking for models and datasets, as well as enhance the visual concealment of the backdoor trigger.

%========================================================================================================================

\ifCLASSOPTIONcaptionsoff
  \newpage
\fi



\bibliographystyle{IEEEtran}
\bibliography{IEEEabrv}

% H.~Kopka and P.~W. Daly, \emph{A Guide to {\LaTeX}}, 3rd~ed.\hskip 1em plus

% \end{thebibliography}

\begin{IEEEbiography}[{\includegraphics[width=1in,height=1.25in,clip,keepaspectratio]{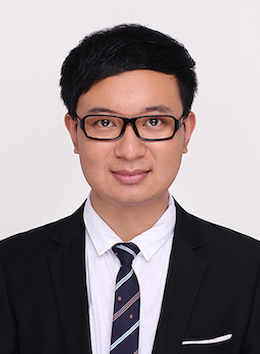}}]{Xiaolei Liu} is an associate research fellow in Institute of Computer Application, China Academy of Engineering Physics. He received the Ph.D. degree and M.S. degree in software engineering from the University of Electronic and Technology of China (UESTC). His research interests include System Security, AI Security and Privacy Protection.
\end{IEEEbiography}

\begin{IEEEbiography}[{\includegraphics[width=1in,height=1.25in,clip,keepaspectratio]{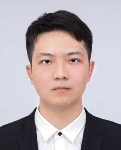}}]{Ming Yi} is a M.S. student in Computer Application at the China Academy of Engineering Physics, Mianyang, China. He received the B.S. degree in Mechanical Design, Manufacturing and Automation from Wuhan University in Wuhan, China. His research interests are focused on AI security.
\end{IEEEbiography}

\begin{IEEEbiography}[{\includegraphics[width=1in,height=1.25in,clip,keepaspectratio]{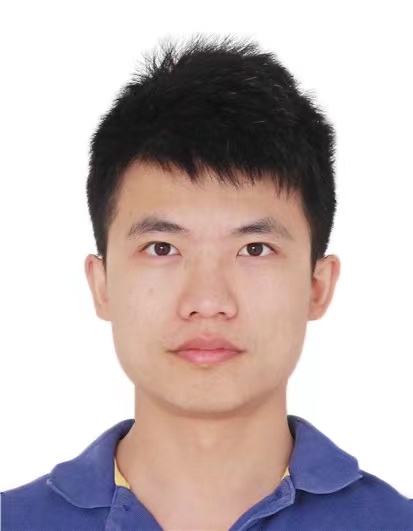}}]{Kangyi Ding} is an assistant research fellow in Institute of Computer Application,  China Academy of Engineering Physics. He received the Ph.D. degree, B.S. degree and M.S. degree in University of Electronic and Technology of China (UESTC). His research interests include AI security, adversarial sample and adversarial transferability.
\end{IEEEbiography}

\begin{IEEEbiography}[{\includegraphics[width=1in,height=1.25in,clip,keepaspectratio]{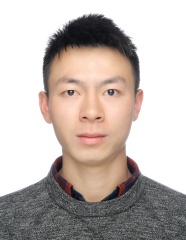}}]{Bangzhou Xin} received his PhD degree in School of CyberScience, University of Science and Technology of China. His research interests include data privacy and machine learning.
\end{IEEEbiography}

\begin{IEEEbiography}[{\includegraphics[width=1in,height=1.25in,clip,keepaspectratio]{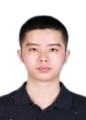}}]{Yixiao Xu} is a Ph.D. student in Cyberspace Security at the Beijing University of Posts and Telecommunications. He received a Master's Degree in Computer Application at the China Academy of Engineering Physics, Mianyang, China. His research interests are focused on AI security, Privacy Protection, and Deep Learning.
\end{IEEEbiography}

\begin{IEEEbiography}[{\includegraphics[width=1in,height=1.25in,clip,keepaspectratio]{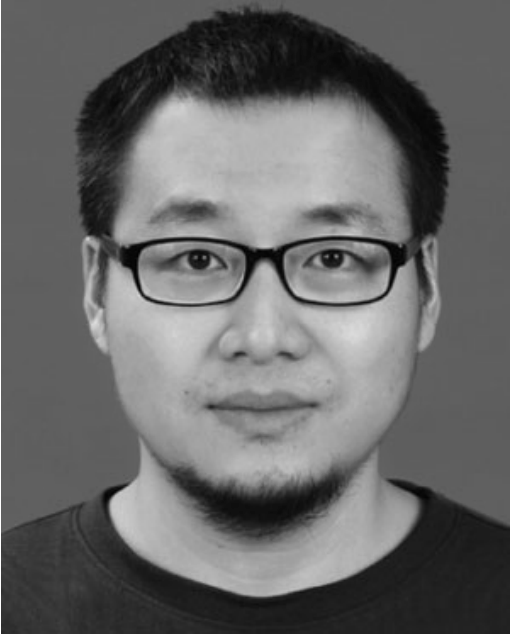}}]{Yan Li}  received the BS degree in information engineering from Xi’an Jiaotong University, Xi’an, China, in 2010, and the PhD degree in computer science from the University of Virginia, Charlottesville, in 2019. He is currently an associate professor with the School of Cyber Science and Engineering, Xi’an Jiaotong University, Xi’an, China. He worked as a postdoctoral researcher with Senseable City Lab, Massachusetts Institute of Technology, Cambridge, from 2020 to 2021. His research interests include data-driven cyber-physical systems, Big Data analytics, and mobile computer networks. He was the Best Transactions Paper Awardee of IEEE Transactions on Intelligent Transportation Systems and the Best-in-Session-Presentation Awardee of INFOCOM’2017.
\end{IEEEbiography}

\newpage

\begin{IEEEbiography}[{\includegraphics[width=1in,height=1.25in,clip,keepaspectratio]{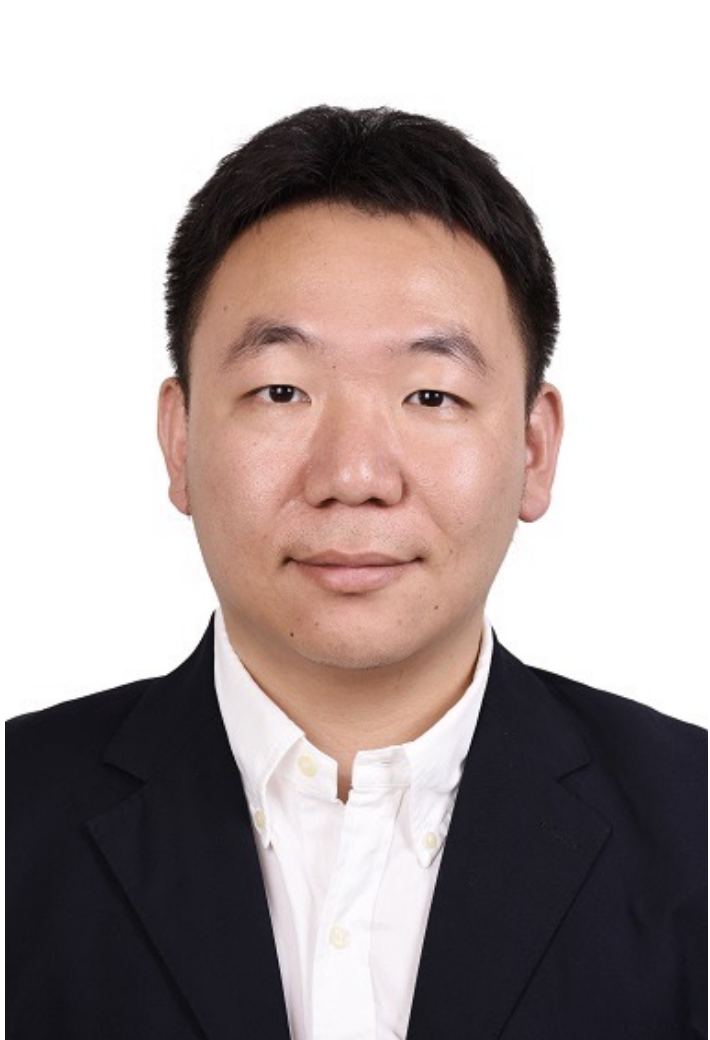}}]{Chao Shen} (Senior Member, IEEE) received the B.S. degree in information engineering and Ph.D. degree in system engineering from the School of Electronic and Information, Xi’an Jiaotong University, Xian, China, in 2007 and 2014, respectively. He is currently a Professor with the School of Electronic and Information Engineering, Xi’an Jiaotong University. He is the Associate Dean of School of Cyber Security, Xi’an Jiaotong University. He is also with the Ministry of Education Key Lab for Intelligent Networks and Network Security. From 2011 to 2013, he was a Research Scholar with Carnegie Mellon University, Pittsburgh, PA, USA. His research interests include network security, human–computer interaction, insider detection, and behavioral biometrics.
\end{IEEEbiography}


% % insert where needed to balance the two columns on the last page with
% %\newpage



%\vfill


\end{document}